# Effective diffusion rates and cross-correlation analysis of "acid growth" data

Mariusz Pietruszka[*] and Aleksandra Haduch-Sendecka

Laboratory of Plant Physiology, Faculty of Biology and Environment Protection, University of Silesia, Katowice, Poland

20.06.2015

[*]Author for correspondence:

Mariusz Pietruszka

University of Silesia, Faculty of Biology and Environment Protection,

Laboratory of Plant Physiology, Jagiellońska 28, PL-40032 Katowice, Poland

Tel: +48-32-2009453, Fax: +48-32-2009361 E-mail: mariusz.pietruszka@us.edu.pl

**Abstract** We investigated the growth-temperature relationship in plants using a quantitative perspective of a recently derived growth functional. We showed that auxin-induced growth is achieved by the diffusion rate, which is almost constant or slowly ascending in temperature, while the diffusion rate of fusicoccin-induced growth increases monotonically with temperature for the entire temperature range [0 – 45] $^oC$, though for some concentrations of IAA "super-diffusion" takes place for unperturbed growth. We also calculated cross-correlations, and the derivative of cross-correlations, for elongation growth (rate) and pH as a function of time delay (lag) parameterized by temperature, for artificial pond water (APW) control conditions (endogenous growth) and exogenous indole-3-acetic acid (IAA) and fusicoccin (FC) introduced into the medium. Dimensionality analysis revealed that discontinuities in cross-correlation derivative correspond to $H^+$ ion kinetics, attaining definite numerical values, approximately proportional to the (logarithm of) proton secretion rates (or relative buffer acidification). Furthermore, three kind of experiments were compared: for abraded coleoptiles, coleoptile segments and intact growing seedlings. From cross-correlation analysis it was found that the timing of IAA and FC-induced proton secretion and growth matches well. Unambiguous results, concerning fundamental conditions of the acid growth hypothesis, were obtained by cross- and auto-correlation analysis: (1) For abraded coleoptiles, because of the lowering of the cuticule potential barrier, auxin-induced cell wall pH decreases simultaneously with the change in growth rate; no advancement or retardation of pH (proton efflux rate) or growth rate takes place (2) Exogenous protons are able to substitute for auxin causing wall loosening and growth (3) Although the underlying molecular mechanisms vastly differ, a potent stimulator of proton secretion, the fungal toxin FC, promotes growth similar to auxin, however of much elevated intensity; as for auxin – no advancement or retardation takes place.

**Introduction**

Plants evolve within the universal constraints imposed by the plant cell wall, dynamically equilibrating the turgor pressure inside the wall (Lintilhac 2014). At the lowest level, the description of cell/plant organ evolution may be expressed in terms of the biophysics and mechanics of the cell wall during growth. Cell extension growth in turgid plant cells/organs is brought about by loosening of the structure of the growth-restraining cell walls, resulting in the relaxation of wall tension and concomitant water uptake (Schopfer 2001). However, the biochemical mechanism of this wall loosening reaction has not yet been fully elucidated. Numerous proteins were recognized as catalysts, in particular cell-wall polysaccharides or expansin family that cause stretch-dependent creep in acidified cell walls by breaking intermolecular non-covalent bonds (Cosgrove 1999, 2000). Primary wall extension growth ("diffusive growth") is fundamental to plant morphogenesis and the evolution of shape. Permanent volume increase must be accompanied by some kind of stress relaxation, otherwise the enlarged cells would tend to shrink to the original size by elastic interactions. Therefore a viscoelastic stress relaxation response is required (we cite after Lintilhac (2014): Metraux and Taiz, 1978; Dorrington 1980; Taiz 1984; also Haduch-Sendecka et al. 2014, Eqs 6.1 – 6.3 and the comment therein). Wei and Lintilhac (2003, 2007) have suggested a different approach to model stress relaxation behaviour. While agreeing with the fact that the source of the tensile stress is turgor pressure, they implied that stress relaxation in plant cell walls (at critical pressure) should be treated as a binary switch, a mechanism which may be appropriate in short-term growth processes (Zajdel et al. 2015), caused by low amplitude, high frequency (osmotic) pressure fluctuations, like in pollen tubes. In recent paper by Pietruszka and Haduch-Sendecka (2015) a solitary frequency $f_0 \approx 0.066$ Hz was determined by the detrended Fast Fourier Transform (FFT) of the wall pressure power spectrum, which reveals strict periodicity in turgor pressure of growing lily pollen tubes – data measured in pressure probe experiment by Benkert et al. (1997) and reanalyzed by Zonia and Munnik (2011). A distinct proposal, also leading to quasi-discrete energy levels resembling a binary switch, was put forward independently by Pietruszka (2013a), in the case of periodic growth of pollen tubes, where asynchronous growth dynamics was achieved through an anharmonic potential at constant turgor pressure condition.

In spite of the extensive efforts to explain the effect of temperature response of plants, the subject seems to be insufficiently appreciated and, in this context, the studies of plant cell/organ in the current literature are rarely reported. Usually, focus is put on growth/development/elongation as a function of temperature, but the plots of such temperature dependence are infrequently presented. Only a few papers in which temperature response is treated as a key issue can be mentioned (see next paragraph). It was suggested that cell growth, especially cell elongation, has a high $Q_{10}$ factor (which is a measure of the rate of change of a biological or chemical system as a consequence of increasing the temperature by 10 °C), which indicates that this is a chemically rather than physically controlled phenomenon (Went 1953). In the above context, we will show that physical (temperature) constraints act through chemical reactions to direct growth.

Temperature is one of the most important factors that determines plant growth, development and yield (Yan and Hunt 1999). It is intelligible that the accurate indication of plant temperature response is a prerequisite to successful crop management. All biological processes respond to temperature, and several models have been proposed. (a) A linear model which is convenient when the temperature does not approach or exceed the optimum temperatutre (Summerfield and Roberts 1987). (b) A bilinear model describing separately sub- and supra-optimum temperatures (Olsen et al. 1993) where the derivations may not be always meaningful and the estimates could be inexact and divergent. (c) A multilinear, multiple parameter (of usually highly empirical origin) model (Coelho and Dale 1980) adopted by crop system simulation packages. (d) Smooth exponential and polynomial models highly inaccurate at both low and high temperature ends (Cross and Zuber 1972; Yan and Wallace 1996, 1998), to name a few. The standard density function of beta-distribution was proposed by Yan and Hunt (1999), using only three cardinal numbers. This approach delivered a universal scaling function, although parameters were not of the kind of 'fine-tuning' parameters somehow rooted in any accompanying microscopic model. The short term temperature response of coleoptile and hypocotyl elongation growth has also been considered by Lewicka and Pietruszka (2008) based on the Central Limit Theorem for several species (barley, wheat, millet, bean, and pumpkin). For our future use it is important to note that in this model the Gauss function was employed.

Empirical elongation/growth studies usually include the notion of temperature implicitly, while some of them intentionally considered the effect of temperature on elongation growth (Karcz and Burdach 2007), which in this work exhibited a clear maximum at 30 °C in maize, and an upward shift of the maximum in the presence of indole-3-acetic acid (IAA) and fusicoccin (FC). High (supra-optimal) temperatures also promoted auxin-mediated hypocotyl elongation in *Arabidopsis* (Gray et al. 1998), who showed that *Arabidopsis* seedlings grown in the light at high temperature of 29 °C exhibit dramatic hypocotyl's elongation compared with seedlings grown at 20 °C. These results strongly supported the contention that growth at high temperature promoted increase of auxin levels and that endogenous auxin promoted cell elongation in intact plants. For the record, beyond the above mentioned papers, we recall recent surveys on growth – temperature relation in plants: Pietruszka et al. 2006; Pietruszka et al. 2007; Pietruszka 2009; Lewicka and Pietruszka 2010, where the preliminary phenomenological models were supported by the authors' performed experiments. The latter were lacking the molecular context, which we supplement in this work.

Many models were developed in the growth area. We name only few of them. A hormone model of primary root growth where the wall extensibility is determined by the concentration of an unspecified by the authors wall enzyme, whose production and degradation are assumed to be controlled by auxin and cytokinin, was proposed by Chavarria-Krauser et al. (2005). More recently Pietruszka (2012) formulated a biosynthesis/inactivation model for enzymatic wall loosening factors or non-enzymatically mediated cell evolution based on the Lockhart/Ortega type of equation. In this work the physiology and biochemistry of the growth process were related by analytical equations acquiring very high fidelity factors ($R^2 \approx 0.9998$, regression $P < 0.0001$) with the empirical data. Also, in the same context of biosynthesis, biological growth as a resultant effect of three forms of energy (mechanical, thermal and chemical) and their individual couplings, was summarized in the form of an elegant theoretical framework by Barbacci et al. (2013). In their description biological growth was the resulting effect of three forms of energy and their couplings (denoted M/T, M/C and T/C with M for Mechanical, T for Thermal and C for Chemical). For each energy, each couple of intensive and extensive variables was linked by one component of Tisza's matrix. This derivation, although sophisticated, requires many parameters (13) and externally controlled turgor pressure *P* and temperature $T$ to retrieve, as an example, the data extracted from the Proseus and Boyer (2008) experiment (see Fig. 5 in Barbacci et al. 2013).

Fast growth of plants requires optimal temperature (e.g. Lewicka and Pietruszka 2006). Below or above this temperature the growth of plant cells and organs is slowed down (ibid., Fig. 3). The latter statement implies that an optimum must exist at the crossover temperature region if one observes slow growth at both high and low temperature ends. At this temperature, it may be presumed, that at least one major factor of the wall-extension governing parameters (couplings) must change intensely. For this approach purposes, we may call it the effective "diffusion rate" $k_2$ (there may be more such coefficients, forming a multi – Gaussian peak spectrum for each coupling $k_i$ in the $k$ – parameters hyper-plane). For practical, e.g. agricultural, use this parameter(s) value(s) is/are of major significance. While extending the model, additional relevant coupling strengths ($k_i$, i ≥ 2) must be added, see Pietruszka (2012), Eq. (18) and the comment therein. We believe that proposal included in this paper and in Zajdel et al. (2015) may form a new base for crop system simulation packages through the included software.

In this work, we consider plant enlargement in the context of reactions linking the process to cell wall biosynthesis. Excellent overviews have been given by Kutschera (2001), Fry (2004), Cosgrove (2005), Boyer (2009) and quite recently by Lintilhac (2014) and Braidwood et al. (2014). However, the steps in wall assembly and the specific chemistry controlling rates of enlargement are still lacking the analytic background. Here we partially covered this broad problem by considering the temperature dependence of wall biosynthesis related factors.

**Material and methods**
**Material** The manuscript is mostly built upon the experimental data which originate from Karcz and Burdach (2007) paper. Briefly, these experiments were carried out with 10 mm long segments cut from 4-day-old maize coleoptiles of maize (*Zea mays* L.) 3 mm below the tip, in usual growth conditions. The experiments were carried out within seven hours each, with measurements taken every 15 min. These raw individual values were retrieved by us with GetData Graph digitizer and collected in SI Tables 1 – 3 for re-analysis. The experimental data obtained by this routine were used in the fitting procedure interrelating the elongation growth data of coleoptile segments and growth functional at constant temperature (Fitexex program, Python code, Zajdel et al. 2015). The definite problems in the datasets used for the analysis are illuminated and discussed later.

Partially we based our analysis on the data published by Lüthen et al. (1990) and Peters et al. (1998), where the abrasion technique was used. For the cross-correlation analysis we digitalized Figs 4 – 5 and Fig. 3, respectively, presented therein.

Finally, for comparison, we present results of our own (48 hours) measurements, performed with the help of a CCD camera, on intact growing maize seedlings (SI Fig. 4). Seeds of maize were grown in the dark at 27 °C; 4-day-old seedlings of a length of about 2.5 cm were chosen for experiment. The experiment was carried out in both chambers simultaneously for APW (first chamber for control) and for the changing growth factor IAA and FC with concentrations from the interval: $0.5 \cdot 10^{-7} - 10^{-5}$ M introduced to the second chamber from the beginning of experiment. The fluid volume in both chambers equaled 30 ml, with 3 seedlings in each chamber. The seedlings of maize were grown in dim green light. A constant temperature was maintained at about 25 °C, and pH, changed by the soaked part of the seedlings and the root system, was measured by two pH-meters (in each chamber independently), type pH/ion meter CPI-501. The images were recorded by Hama Webcam AC-150 every 30 min. From length and time measurements, the relative elongation of marked coleoptile segments (initially 1 cm long fragments indicated by ink spots) was calculated using the formula $(l_f - l_i)/l_i$, where $l_i$ is the initial length and $l_f$ is the final. This method allowed for simultaneous measurements of growth and $H^+$ efflux. The software OriginPro 8.5.1 (Microcal) was utilized to perform calculations and create graphs in all cases.

**Methods**

*Relative elongation growth formula*

For mathematical analysis we used the data collected in SI Tables 1 – 3 (presented in SI Figs 1 – 3) for the double-exponent growth functional derived in Pietruszka (2012, 2013b) and further elaborated in detail for practical use in Zajdel et al. (2015) for the relative (volumetric) growth rate

$$\frac{V_T(t) - V_0}{V_0} = At + B + C(T) e^{-e^{-D(T)(t-t_i)}} \tag{1}$$

where *A*, *B*, *C* and *D* are positively defined coefficients constant in time. The coefficients $C = C(T)$ and $D = D(T) = k_2(T) = 1/\tau_2$ are, in principle, temperature (*T*) dependent. A particular time $t = t_i$ stands for the inflection point (corresponding to the maximum in growth rate) of the sigmoid-like growth curve (Pietruszka 2012, Fig. 8A). In present analysis the effective diffusion rate which involves net transport to the wall $k_2(T)$ is our main interest, since it can be directly inferred from the fits. Its value depends on dimensionality like $[D]=1/s$, and therefore can be compared between diverse species in different experimental conditions. Note, that coefficient $k_2 = D$ (diffusion rate) should not be confused with the diffusion *constant D* defined in the I$^{st}$ Fick's law.

To prepare temperature-dependent data for further analysis, the additional (intermediate) data points were obtained by linear interpolation (moving average) – SI Table 4 – 6.

We also use for analysis the probability density function of the beta distribution (also called the Euler integral of the first kind (Polyanin and Chernoutsan 2011)), for the interval $0 \leq x \leq 1$, and shape parameters α, β > 0. Beta function is a power function of the variable x and of its reflection (1 − x)

$$f(x;\alpha,\beta) = \text{const} \cdot x^{\alpha-1}(1-x)^{\beta-1} = \frac{x^{\alpha-1}(1-x)^{\beta-1}}{\int_0^1 u^{\alpha-1}(1-u)^{\beta-1}du} = \frac{1}{B(\alpha,\beta)} x^{\alpha-1}(1-x)^{\beta-1} \qquad (2)$$

and has a normalization constant $B = B(\alpha, \beta)$. The use of beta distribution was already suggested by Yan and Hunt (1999) for temperature dependent plant growth. The coexistence curve (Fig. 3, ibid.) presented by the authors strongly supports the use of beta function also in this study. The results of the fitting procedure for the growth process amplitude $C = C(T)$, are related to the beta function (2) in the following way

$$C(T;\alpha,\beta) = [c_T \cdot T^{\alpha-1}(1-T)^{\beta-1}]_{pH} \qquad (3)$$

where $c_T$ is a constant. The rescaled value of temperature $T$ belongs to [0, 1] interval. A universal scaling can be expected in a form of a coexistence curve (Stanley 1971, Fig. 1.8; Pietruszka 2015).

Fitting Eq. (1) to the experimental data provides initial parameters $A$, $B$, $C$, $D$ and $t_i$, under a condition that the proper unit scaling was done. We have already delivered (Zajdel et al. 2015) the physiological explanation of these parameters, though the interpretation of parameter $C$ was still lacking. Then, we could have only assigned a process "amplitude" meaning to coefficient $C$. Now, in the context of present analysis, it seems, that parameter $C = C(T)$, if dependent on temperature, is biologically meaningful. It can be successfully described in temperature context by the Euler beta function, or equivalently by pH changes in the apoplast (Pietruszka 2015), leading to proton extrusion into the wall compartment and subsequently – in accord with "acid growth hypothesis" (Hager 2003) – viscoplastic wall extension. In this way, temperature, connected with the rate of chemical reactions ($Q_{10}$ factor), or pH – attributed to plasma membrane PM $H^+$- activity, can be introduced into the growth functional Eq. (1). Eq. (1) faithfully reproduces experimental results (with determination coefficient attaining in the majority of cases extremely high value, especially when biological experiments are taken into account: $R^2 \sim 0.9998$ (Zajdel et al. 2015)). In conclusion, $C \equiv C(T; \alpha, \beta) \sim C$ (pH), and $D$ recognized earlier as "diffusion rate", reflects the rate of particles flow. Coefficient $A \sim (P - Y)$ and parameter $B \geq 0$ has – as yet – no interpretation. Henceforth we will consider only $C(T)$ and $k_2 = D(T)$ coefficients.

The latter paragraph needs an additional comment. According to the second Fick's law diffusion rate gives the relation between the temporal concentration derivative (accumulation) and the spatial concentration distribution (gradient) of a solute species in a given medium/solvent (e.g. water), is denoted by $D$ and has units [m$^2$/s]. The effective diffusion rate $k_2$ as referred to in this paper is by its nature the reciprocal of the characteristic process time constant $\tau_2$ which has unit [s] (and so $k_2$ is in [1/s] units). According to the model constitutive equation, Eq. (1), $D$ is exactly $1/\tau_2$. Given that the origin of $D$ is semi-phenomenological, although rooted in processes ultimately governed by thermodynamic processes such as diffusion, and considering the fact that the relative volumetric change (according to Eq. (1)) is very complex in origin, we proposed to denote $k_2$ ($= D$) as an effective diffusion rate (see Zajdel et al. 2015 for further elucidation).

*Cross-correlations*

In signal processing, cross-correlation is a measure of similarity of two waveforms as a function of a time-lag applied to one of them. This is also known as a sliding dot product or sliding inner-product. For continuous functions $f$ and $g$, the cross-correlation is defined by the integral

$$(f * g)(\tau) \equiv \int_{-\infty}^{\infty} f^*(t) g(t+\tau) dt \tag{4}$$

where $f^*$ denotes the complex conjugate of $f$ and $\tau$ is the time lag. Note, that the cross-correlation is maximum at a lag equal to the time delay (maximum located at lag equal zero means no time delay). Similarly, for discrete functions (like the usually analysed experimental data points), the cross-correlation is defined as:

$$(f * g)[n] \equiv \sum_{m=-\infty}^{\infty} f^*[m] g[m+n] \tag{5}$$

which definition is utilized in this work (here: $f^* = f$). Auto-correlation is obtained if $f$ equals $g$. Cross-correlation derivative (over time delay $\tau$) is defined as follows (see also Appendix)

$$\frac{d}{d\tau}(f * g)(\tau) \equiv \frac{d}{d\tau}\left[\int_{-\infty}^{\infty} f^*(t) g(t+\tau) dt\right] \tag{6}$$

**Results and discussion**

For discussion purposes we omit details of the derivation of Eq. (1), which can be found elsewhere (Pietruszka 2012; Zajdel et al. 2015). Instead, we cover the subsequent thought leaps leading to the temperature dependent $C$ and $D$ coefficients (fit parameters) in Eq. (1).

*Effective diffusion rates*

First, let us recall (Pietruszka 2012) the first-order differential equation for $n(t)$

$$\frac{dn(t)}{dt} = k_1 - k_2 n(t) \tag{7}$$

which is an actual concentration of WLF (Wall Loosening Factor) which solution was inserted into the truncated form (for constant turgor $P$) of the Ortega (1985) equation. In Eq. (7) the most interesting for our problem is the diffusion rate $k_2$, which is the rate of change of WLF concentration in the cell wall at a given temperature. The coefficient $k_1$ which originates from the external pool in the model calculations (it may be also interpreted as a 'biosynthesis' coefficient), is incorporated into the remaining coefficients of Eq. (1).

With the usual assumption that both cell wall extension and water uptake must occur concomitantly, and the overpressure (by tissue impact) dependence of the yield threshold $Y = Y[n(t)]$ - in the first approximation - may be neglected, we arrived at the solution

$$V(t) = V_0 \exp[-\frac{e^{-k_2 t} n_0 (P-Y)}{k_2}] \qquad (8)$$

where $n_0 = n(t=0)$. We recall, that in general the kinetic coefficient $k_2 = k_2(T)$ is temperature dependent (Fig. 7 in Pietruszka 2010). Hence, relevant but not mutually interacting coupling strengths ($k$) must be taken into account while extending the model. In the extended form the equation for the relative (logarithmic) growth reads

$$\frac{1}{V(t)}\frac{dV(t)}{dt} = \Phi_0 \sum_{i=1}^{n} \frac{1}{w_i} x_i(t)(P(t) - Y[n_i(t)]) + \frac{1}{\varepsilon}\frac{dP(t)}{dt} \qquad (9)$$

where the completeness relation $\sum_{i=1}^{n} w_i = 1$ formulates the constraint for positive $w_i$ – weight of $i^{th}$ constituent (WLF, which may be of endogenous or exogenous origin, see Pietruszka 2012 for further discussion); $x_i$ denotes $i^{th}$ constituent concentration. In Eq. (9) $\Phi_0$ is the Lockhart (1965) constant responsible for cell wall plastic extensibility, while $\varepsilon$ is the elastic constant (Ortega 1985). Note, the second order correction for the yield stress $Y$ naturally enters Eq. (9) *via* functional dependence. As it was already noticed (Pietruszka 2012) the model does handle *in vivo* and *in vitro* activities in similar terms.

Next, we recall that the plots presented in Fig. 2 in Pietruszka (2012) exhibit pronounced changes with respect to the 'coupling constant' strength (effective diffusion rate $k_2$), in contrast to lesser reactions caused by turgor pressure change. The interpretation from the analytic expression (ibid.) follows that volumetric extension to be effective must be preceded by pressure induced relaxation processes in the cell wall due to WLFs interaction with the wall constituting polymers (Schopfer 2008; Geitmann and Ortega 2009) – otherwise growth is less successful since the wall is more 'rigid' and not susceptible to the pressure changes.

We also bear in mind that the model involves biosynthesis of WLFs in the cell at a steady rate $k_1$ and partial inactivation of such created WLFs at a rate $k_2$. Note, that $k_2$ (and/or $k_i$, $i \geq 2$ in case of Eq. (9)) can be in general temperature dependent, as it is shown in Fig. 7 in Pietruszka (2010). The advantage of this proposal (Pietruszka 2012; Zajdel et al. 2015) is that by preserving appropriate experimental conditions during measurements of $V(t)$, or more precisely $\frac{V_T(t) - V_0}{V_0}$ *versus* time, the obtained set of $k$ values may return important information not only about the interaction strengths on the molecular level. The regulatory role of endogenous WLF chemical activity is shown in Figs 1 and 2, especially pronounced in Fig. 2A, where a sharp Gaussian peak appears. We need to point out that the obtained diffusion rate $k_2$ cannot be directly linked to the growth rate, since the volumetric increase in volume can be also built-in into the coefficient $C = C(T)$, which serves as nonlinear, temperature-dependent growth amplitude in Eq. (1). A good example can be drawn by comparison of the results in SI Table 1P (Zajdel et al. 2015), where the diffusion rate $k_2$ (parameter $D$) is slightly different for the hypocotyls grown in the dark/light conditions. The over tenfold increase in length is incorporated into parameter $C$, Fig. 3A – C (ibid.). Note, that the greater $k_2$ the more substantial decrease of the initial concentration $n_0$, and the quicker decrease of the actual 'growth factor'.

Based on the results from this paper, it was shown that '$k_2$ factor' decisively influences cell/organ volume. By adding a biochemical substance, which causes similar effect (as WLF), one should observe shifts of the peak in Fig. 2A. Hydroxyl radicals (OH)⁻ are capable of unspecifically cleaving cell wall polysaccharides in a site specific reaction (Schopfer 2001). Cell wall loosening underlying the elongation growth of plant organs is controlled by apoplastically produced OH⁻ "attacking load-bearing cell wall matrix polymers" (ibid.).

In the above scenario, molecular factors can be exemplified by:

- the dependence of extension on concentrations of ascorbate/$H_2O_2$ and $Cu^{2+}$ or $Fe^{2+}$ (used for generating OH⁻ in isolated cell walls of maize coleoptiles (Fig. 3, Schopfer 2001)),
- the dependence of extension on pH (Fig. 4, ibid.),
- inhibition of auxin-induced elongation growth by Mn-based chemicals (Fig. 9, ibid.).

Implications of our proposal are also supported by a clearly visible shift in Porter and Gawith (1999) study, Fig. 1.

For further discussion we call to mind the results of Lewicka and Pietruszka (2006). Especially, we draw reader's attention to comparison of Fig. 6 (ibid.) with three characteristic phases (crystalline, semi-liquid and liquid) to the endogenous growth amplitude coefficient $C$ from Eq. (1), presented in Fig. 1A in the present work. From the biomechanical point of view cell membranes are equipped with ionic pumps and channels, water channels and ligand receptors (Berg et al. 2002). The optimal phase of endogenous growth starts to occur in the semi-liquid phase, presumably corresponding to the peak of the Gauss curve located at the phase boundary at about 16 °C in the present work (Fig. 2A) and a "jump" above 16 °C in Fig. 1A. Then, the endogenous auxin activates $H^+$- ATPase, acidification of cell walls and their loosening. Simultaneously, the $K^+$ ions in-flow takes place, through the reduction of the water potential filling up the plant interior, pulling behind water molecules. At this temperature end we encounter a kind of a 'phase transition' from a crystalline to semi-liquid to phase. Low temperatures cause membrane depolarization and $K^+$ loss, water efflux and growth inhibition. Furthermore, at 0 °C the reservoir of liquid water becomes nearly empty for the sake of crystallisation into the ice phase. At high temperatures, another transition from a semi-liquid to a liquid phase occurs (arrow pointing at a local maximum at 38 °C, Fig. 2a), causing the malfunction of ionic and water channels as well as ionic pumps (and hence the fits for $k_2$ become inexact, which can be observed at temperatures above 40 °C in Fig. 2a). In effect, we deal with the ionic leakage, a secondary water stress and consequently cessation of growth. Moreover, the auxin receptor proteins change conformation and functionality, causing growth deceleration and termination. The described processes are reflected in Figs 1A – C and Figs 2A – C in this paper. Further research should address how the magnitude of the related biophysical variables and specific chemical reactions responsible for the optimum growth are affected by temperature (and pH) within the cell. It seems, that at sufficient water supply conditions the growth rate of plant cells or organs can be maintained optimal either by change of the environmental temperature or by fine-tuning the system through pH-dependent biochemical reactions (Pietruszka 2015). The difficulty lies in finding the chemical potential-dependent essential reaction(s) responsible for modification of *C* and *D*. Determination of such reaction(s) can be of great significance for the food production problem in areas of sub- and supra-temperatures, where pH-induced biochemical factors can compensate for non optimal growth temperatures of the environment.

Besides, descriptions of experimental datasets (SI Figure 1 – 3 and SI Table 4 – 6 ) concluded in coefficients *C* and *D* were even better than those that can be provided by the generalised logistic (6 parameters) Richards function (1959), the most flexible of the classical growth equations. This is mostly due to the fact that (beyond of the least number of parameters), the parameters used in the double exponential function proposed by us (Pietruszka 2012) possess real biochemical and biophysical underpinning. Besides, it seems, that parameter $C = C(T)$, if dependent on temperature, is biologically meaningful, when described by Euler beta function (Yan and Hunt 1999). Coefficient *C* can also designate pH changes in the apoplast and hence – proton extrusion into the wall (Pietruszka 2015).

*Cross-correlation analysis*

Some crucial arguments against the acid growth theory of auxin action (Kutschera and Schopfer 1985) have been reinvestigated by simultaneous measurements of proton fluxes and growth of *Zea mays* L. coleoptiles by Lüthen et al. (1990). Among others, it was found that (a) the timing of auxin and fusicoccin-induced (FC) proton secretion and growth matches well and (b) the equilibrium external pH in the presence of IAA and FC are lower than previously recorded and below the so-called "threshold-pH". It was concluded that the acid-growth-theory correctly describes incidents taking place in the early phases of auxin-induced growth. This subject was undertaken next by us and the results are summarized in Table 1 and Figs 3 – 10. To save space, the description of our results are partially located in figures captions.

Cross-correlations of pH and elongation growth as a function of time delay $\tau$, Eq. (5), for APW (endogenous growth), exogenous IAA and FC is shown in Fig. 3. A similar plot parameterized by temperature is presented in Fig. 4. The analysis was based on the raw data presented in Figs 1 – 5 in Karcz and Burdach (2007). The time retardation (advancement) of the maximum with respect to zero time delay is clearly visible (see also SI Table 7 for the obtained values).

Cross-correlations derivative of elongation growth and pH as a function of time lag $\tau$, Eq. (6), parameterized by temperature, for APW (endogenous growth), exogenous IAA and FC are shown in Figs 5 – 6. The discontinuities (representing the relative buffer capacity acidification, approximately proportional to proton efflux rate) in the cross-correlation derivative at $\tau = 0$ correspond to $H^+$ ions activity for maize coleoptile segments (see also SI Table 8 for the obtained values).

There are some general problems with the Karcz and Burdach (2007) data, as they did not use abraded coleoptiles like in Lüthen et al. (1990). Instead, they did perfuse the coleoptile cylinders with solution. That appears to work to a certain degree as indicated by the ± nominal pH drops. It may well be that the responses in this system are a bit more sluggish than in abraded coleoptiles. This apparently might affect the meaningfulness of any cross correlation analysis. This seems not quite to be the case. The abraded coleoptiles, like in Lüthen et al. (1990) or coleoptile segments in Karcz and Burdach (2007) transducer experiments simply provide us with different cross-correlation reactions. The response is more or less delayed, as it is shown in Figs 3 – 4 and Figs 7 – 9. Although such data does not deliver unambiguous claims concerning (time lags in) the acid growth hypothesis, they can probably be useful in analysing the influence of a cuticule potential barrier. This task can presumably be further accomplished by Monte Carlo simulations of thermal protons efflux as a function of the properties of such a barrier (Pietruszka and Konefał 2015). This is due to the fact that the height of a barrier, that is dependent on cuticule properties, should be reflected in time delays obtained in cross-correlation analysis. The latter remark should hold at least for ideal-experiment model analysis.

Another problem with the Karcz and Burdach (2007) data is the fact that auxin and FC were added very early in the pH-drop kinetics. There is a very careful analysis of such pH drops (Peters and Felle, 1991) which very well describes what goes on in a well-performed pH drop experiment. Basically these authors, in order to clarify discrepancies between the earlier Lüthen et al. (1990) and Kutschera and Schopfer (1985) papers, found that the pH of excised coleoptile segments first rises to pH 6.5 (RT – reversal time), and then gradually falls to an AE phase (acid equilibrium of about pH 4.8 in maize), which is achieved about 4 hours after excision. When auxin is added, the pH will drop to 4.2 with a time course well matching the growth response. This pattern is also well visible in Fig. 4 in Karcz and Burdach (2007). However, as they apply the auxin very early (after two hours), they always see the auxin effects on a large background of the still ongoing endogenous pH drop. On the contrary, in Peters and Felle (1991) auxin was added at the acid equilibrium, making the pH drop much more clearly visible.

To resolve these doubts we analysed the experimental results presented in Figs 4 and 5 by Lüthen et al. (1990). The apparently unattractive outcomes, presented in Figs 7 – 8, we accepted with amazement. In Fig. 7 cross-correlation of growth rate and proton efflux rate is calculated for IAA and FC-induced growth. First, we noted that the timings for each plot matches well and the cross-correlation intensity for FC is several times stronger than for IAA. No time delay is observed. Second, the triangular shape of both curves and the location of both maxima at zero lag brings to mind the definition of autocorrelation (the cross-correlation of a signal with itself at different points in time, Eq. (6) for $f = g$), which is the most striking feature of this – otherwise known – result. It means that cell wall pH and growth rate are strictly co-regulated (strongly correlated) in growing shoot tissue. This observation led us to perform autocorrelation analysis presented in Fig. 8. Comparison of autocorrelation plots in Fig. 8 A with B for auxin-induced growth rate and proton efflux rate and Fig. 8 C with D for fusicoccin, revealed the remarkable similarity of both pairs of plots (A – B) and (C – D). It means, that primary wall growth rate and proton efflux rate can be identified. This result, among others, strongly support the foundations accepted for the derivation of equation of state (EoS) by Pietruszka (2015).

The latter result, valid at least from the mathematical point of view, has been obtained for abraded coleoptiles by Lüthen et al. (1990) – the epidermal cuticule is a strong barrier for protons. Departures from this ideal picture are shown in Fig. 9 where time delays and plot deformations are present as a result of transient changes after addition of IAA.

Finally, for comparison, the results of an 'intact seedling' experiment for maize are shown in Fig. 10 (see also SI Fig. 5). Note, that elongation growth and pH are cross-correlated for auxin and fuscicoccin showing the time lag in majority of cases for FC. For this kind of experiment, the correlation strength is, at least in some cases (concentration $10^{-7}$), more intense for auxin.

**Conclusions**

We considered the temperature-dependent effective diffusion rate with specific applicability to cell wall loosening factors and application of Euler beta distribution to the amplitude $C$ of the growth functional, Eq. (1). It was shown, that the endogenous/exogenous growth amplitude $C = C(T)$ is realistically reproduced by the Euler beta function as a function of temperature, while the temperature-dependent endogenous diffusion rate $D = k_2(T)$ is reasonably represented by Gauss distribution (endogenous auxin) or linear function (exogenous auxin of fusicoccin). In the temperature context, the diffusion rate $k_2 = k_2(T)$ and the amplitude $C$, describing active ($H^+$) transport into the wall, provides a measure of thermal energy efficiency of growth. We showed, that the localisation of endogenous growth maximum is essentially determined by temperature (or equivalently by pH, through plasma membrane $H^+$-ATPase), and that the cumulative action of $C$ and $k_2$ coefficients essentially contribute to growth. It also seems that the localization of the optimum growth is mainly determined by temperature (or pH, Pietruszka 2015) – $C$ coefficient.

The main limitation of our early 'temperature' approach to plant cell/organs growth was that it was not accounting for the important role of the biochemical reactions involved in cell wall building processes. Our present study extends our previous proposals into a new territory. The strongly predicative (temperature dependent) semi-empirical equation (1) permit to fine-tune the leading factors in plant cell/organ growth, which implications may be helpful for climatic impact studies onto plant growth.

Our results also suggest that at least for some special experimental conditions (abraded samples like in Lüthen et al. 1990), the timings of growth and proton efflux match, while the interaction expressed by cross-correlations is much stronger for fusicoccin than for auxin.

The molecular mechanism of fusicoccin and auxin differ: (1) 14-3-3 protein interaction with the ATPase in the case of FC and (2) TIR1 and/or ABP1 binding and an unknown signaling pathway in the auxin case. However, the $H^+$ efflux mechanism by which auxin and fusicoccin cause the promotion of growth may be effectively similar on the level of primary wall tissues. At this lowest (molecular) level, auto-correlation analysis of Lüthen et al. 1990 data led us to the conclusion that the primary wall growth rate and $H^+$- efflux rate coincide

$$\boxed{\text{growth rate} \equiv \text{proton efflux rate}}$$

It seems that investigating "acid growth hypothesis", and resolving mounting controversies, is today impossible using solely biological experiments. It could be explored further using modelling as shown in this work. In the case of this work the *experimentum crucis* for a biological problem belongs – paradoxically – to physics and mathematics.

In conclusion, we believe that model equation (1) we used for calculations (supplemented with an open access computer code) can actually fit a vast number of classical real-world time courses in a number of species and under very different experimental conditions, and, therefore, may perhaps describe the general process of elongation growth in a realistic way.

**Appendix**

The cross-correlation of continuous functions $f$ and $g$ is defined in Eq. (4). By assuming $f \equiv \text{pH}(t)$ and $g \equiv u(t)$ [μm] the cross-correlation derivative (over time delay $\tau$) can be calculated explicitly as follows

$$\frac{d}{d\tau}(pH * u)(\tau) \equiv \frac{d}{d\tau}\left[\int_{-\infty}^{\infty} pH(t)u(t+\tau)dt\right] = \int_{-\infty}^{\infty} pH(t)\underbrace{\frac{d}{d\tau}u(t+\tau)}_{u'}dt = \int_{-\infty}^{\infty} pH(t)u'(t+\tau)dt$$

where $u'$ is a growth rate.

**References**

1. Azuah RT, Kneller LR, Qiu Y, Tregenna-Piggot PLW, Brown CM, Copley JRD, Dimeo RM (2009) DAVE: a comprehensive software suite for the reduction, visualization, and analysis of low energy neutron spectroscopic data. J Res Nat Inst Stand Technol 114: 341

2. Barbacci A, Lahaye M, Magnenet V (2013) Another brick in the cell wall: biosynthesis dependent growth model. PLoS ONE 8: e74400

3. Benkert R, Obermeyer G, Benturp F-W (1997) The turgor pressure of growing lily pollen tubes. Protoplasma 198: 1–8

4. Berg JM, Tymoczko JL, Stryer L (2002) Biochemistry. Freeman & Company

5. Boyer JS (2009) Cell wall biosynthesis and the molecular mechanism of plant enlargement. Funct Plant Biol 36: 383–394

6. Braidwood K, Breuer C, Sugimoto K (2014) Mybody is a cage: mechanisms and modulation of plant cell growth. New Phytol 201: 388–402

7. Chavarria-Krauser A, Jaeger W, Schurr U (2005) Primary root growth: a biophysical model of auxin-related control. Funct Plant Biol 32: 849–862

8. Coelho DT, Dale RF (1980) An energy-crop growth variable and temperature funcion for predicting maize growth and development: planting to silking. Agronomy J 72: 503–510

9. Cosgrove DJ (1999) Enzymes and other agents that enhance cell wall extensibility. Annu Rev Plant Physiol Plant Mol Biol 50: 391–417

10. Cosgrove DJ (2000) Expansive growth of plant cell walls. Plant Physiol Biochem 38: 109–124

11. Cosgrove DJ (2005) Growth of the plant cell wall. Nature Reviews 6: 850–861

12. Cross HZ, Zuber MS (1972) Prediction of flowering dates in maize based on different methods of estimating thermal units. Agronomy J 64: 351–355

13. Dorrington K (1980) The theory of viscoelasticity in biomaterials. In: Vincent JFV, Currey JD (eds) The mechanical properties of biological materials. 34th Symposium of the Society of Experimental Biology

14. Hager A (2003) Role of plasma membrane $H^+$-ATPase in auxin-induced elongation growth: historical and new aspects. J Plant Res 116: 483–505

15. Haduch-Sendecka A, Pietruszka M, Zajdel P (2014) Power spectrum, growth velocities and cross-correlations of longitudinal and transverse oscillations of individual *Nicotiana tabacum* pollen tube. Planta 240: 263–276

16. Fry SC (2004) Plant cell wall metabolism: tracking the careers of wall polymers in living plant cells. New Phytol 161: 641–675

17. Geitmann A, Ortega JK (2009) Mechanics and modelling of plant cell growth. Trends Plant Sci 14: 467–478, Review

18. Gray WM, Ostin A, Sandberg G, Romano CP, Estelle M (1998) High temperature promotes auxin-mediated hypocotyls elongation in *Arabidopsis*. Proc Natl Acad Sci USA 95: 7197–7202

19. Karcz W, Burdach Z (2007) Effect of temperature on growth, proton extursion and membrane potential in maize (*Zea mays* L.) coleoptiles segments. Plant Growth Regul 52: 141–150

20. Kutschera U, Schopfer P (1985) Evidence against the auxin-growth theory of auxin action. Planta 163: 483–493

21. Kutschera U (2001) Stem elongation and cell wall proteins in flowering plants. Plant Biol 3: 466–480

22. Lewicka S, Pietruszka M (2006) Theoretical search for the growth-temperature relationship in plants. Gen Physiol Biophys 25: 125–136

23. Lewicka S, Pietruszka M (2008) Central limit theorem and the short-term temperature response of coleoptiles and hypocotyls elongation growth. Acta Soc Bot Pol 77: 289–292

24. Lewicka S, Pietruszka M (2010) Generalized phenomenological equation of plant growth. Gen Physiol Biophys 29: 95–105

25. Lintilhac P (2014) The problem of morphogenesis: unscripted biophysical control systems in plants. Protoplasma 251: 25–36

26. Lockhart JA (1965) An analysis of irreversible plant cell elongation. J Theor Biol 8: 264–275

27. Lüthen H, Bigdon M, Böttger M (1990) Reexamination of the acid growth theory of auxin action. Plant Physiology 93: 931–939

28. Metraux J, Taiz L (1978) Transverse viscoelastic extension in *Nitella*. 1. Relationship to growth rate. Plant Physiol 61: 135–138

29. Olsen JK, McMahon CR, Hammer GL (1993) Prediction of sweet maize phenology in subtropical environments. Agronomy J 85: 410–415

30. Ortega JKE (1985) Augmented growth equation for cell wall expansion. Plant Physiol 79: 318–320

31. Peters WS, Felle H (1991) Control of Apoplast pH in Corn Coleoptile Segments. I: The Endogenous Regulation of Cell Wall pH. J Plant Physiol 137: 655–661

32. Peters WS, Lüthen H, Böttger M, Felle H (1998) The temporal correlation of changes in apoplast pH and growth rate in maize coleoptile segments. Aust J Plant Physiol 25: 21–25

33. Pietruszka M, Lewicka S, Pazurkiewicz-Kocot K (2006) Thermodynamics of irreversible plant cell growth. Acta Soc Bot Pol 75: 183–190

34. Pietruszka M, Lewicka S, Pazurkiewicz-Kocot K (2007) Temperature and the growth of plant cells. J Plant Growth Regul 26: 15–25

35. Pietruszka M (2009) Self-consistent equation of plant cell growth. Gen Physiol Biophys 28: 340–346

36. Pietruszka M (2010) Exact analytic solutions for a global equation of plant cell growth. J Theor Biol 264: 457–466
37. Pietruszka M (2012) A biosynthesis/inactivation model for enzymatic WLFs or non-enzymatically mediated cell evolution. J Theor Biol 315: 119–127
38. Pietruszka M (2013a) Pressure-induced cell wall instability and growth oscillations in pollen tubes. PLoS ONE 8: e75803
39. Pietruszka M (2013b) Special solutions to the Ortega equation. J Plant Growth Regul 32: 102–107
40. Pietruszka M, Haduch-Sendecka A (2015) Pressure-induced wall thickness variations in multi-layered wall of a pollen tube and Fourier decomposition of growth oscillations. Gen Physiol Biophys 34: 145–156
41. Pietruszka M (2015) pH/T duality – the equation of state for plants. arXiv:1505.00327
42. Pietruszka M, Konefał A (2015) Monte Carlo simulation of proton efflux by a cuticule potential barrier. In preparation
43. Polyanin AD, Chernoutsan AI (2011) A concise handbook of Mathematics, Physics and Engineering Science CRC Press
44. Porter JR, Gawith M (1999) Temperatures and the growth and development of wheat: a review. European J Agronom 10: 23–36
45. Proseus TE, Boyer JS (2008) Calcium pectate chemistry causes growth to be stored in *Chara carolina*: a test of the pectate cycle. Plant, Cell & Environ 31: 11147–55
46. Richards FJ (1959) A Flexible Growth Function for Empirical Use. J Exp Bot 10: 290–300
47. Schopfer P (2001) Hydroxyl radical-induced cell-wall loosening *in vitro* and *in vivo*: implications for the control of elongation growth. Plant J 28: 679–688
48. Schopfer P (2008) Is the loss of stability theory a realistic concept of stress relaxation-mediated cell wall expansion during plant growth? Plant Phys 147: 935–936
49. Stanley HE (1971) Introduction to phase transitions and critical phenomena. Oxford University Press
50. Summerfield RJ, Roberts EH (1987) Effects of illuminance on flowering in long- and short-day grain legumes: a reappraisal and unifying model. In: Atherton JG, ed. Manipulation of flowering. London, Butterworths
51. Taiz L (1984) Plant cell expansion: regulation of cell wall mechanical properties. Ann Rev Physiol 35: 585–657
52. Wei C, Lintilhac P (2003) Loss of stability – a new model for stress relaxation in plant cell walls. J Theor Biol 224: 305–312
53. Wei C, Lintilhac PM (2007) Loss of stability: A new look at the physics of cell wall behavior during plant cell growth. J Theor Biol 283: 113–121

54. Went FW (1953) The effect of temperature on plant growth. Annu Rev Plant Physiol 4: 347–362
55. Yan W, Wallace DH (1996) A model of photo period x temperature interaction effects on plant development. Critical Rev Plant Sci 15: 63–96
56. Yan W, Wallace DH (1998) Simulation and prediction of plant phenology for five crops based on photoperiod by temperature interaction. Annals Bot 81: 705–716
57. Yan W, Hunt LA (1999) An equation for modelling the temperature response of plants using only the cardinal temperatures. Annals Bot 84: 607–614
58. Zajdel P, Haduch-Sendecka A, Pietruszka M (2015) Application of the effective analytical formula of growth functional to quantitative description of growth of plant cells and organs (J Plant Growth Reg, in review). Python code „fitexex – Application of exp(exp()) function for plant physiology" https://github.com/pawelzajdel/fitexex
59. Zonia L, Munnik T (2011) Understanding pollen tube growth: the hydrodynamic model versus the cell wall model. Trends Plant Sci 16: 347–352

**Table caption**

**Table 1** Numerical values of discontinuities obtained in cross-correlation derivative proportional to the relative buffer acidification, expressed as $H^+$ ions activity. See also Figs 5 – 6. Dashed lines – unavailable data.

| Temperature | APW | IAA | FC |
|---|---|---|---|
| 10 °C | 6.33 | 31.41 | 125.40 |
| 25 °C | 219.57 | 385.60 | 383.99 |
| 30 °C | 351.23 | 458.24 | – – – – |
| 35 °C | 261.43 | 422.93 | 427.09 |
| 40 °C | 149.18 | – – – – | 414.95 |

**Figures captions**

**Figure 1 A** Endogennous growth (APW) amplitude coefficient *C* as a function of temperature (solid squares); artificial pond water (APW) incubation medium. Temperature interval [0 – 40] °C corresponds to [0 – 1]. A maximum at 30 °C (Euler beta function fit, β-spline). The calculated data points and error bars – see SI Table 4. The fit parameters calculated from Eq. (2) are indicated in the chart. The goodness of fit: $\chi^2 = 0.0003$ and the determination coefficient $R^2 = 0.94$. **B** Exogennous growth (IAA) amplitude coefficient *C* as a function of temperature (solid squares); auxin (IAA) introduced after 2 hours into the incubation medium. Temperature interval [0 – 45] °C corresponds to [0 – 1]. A maximum at 30 °C (Euler beta function fit, β-spline interpolation). The calculated data points and error bars – see SI Table 5. The fit parameters calculated from Eq. (2) are indicated in the chart. The goodness of fit: $\chi^2 = 0.00018$ and the determination coefficient $R^2 = 0.97$. **C.** Exogennous growth (FC) amplitude coefficient *C* as a function of temperature (solid squares); fusicoccin (FC) introduced after 2 hours into the incubation medium. Temperature interval [0 – 45] °C corresponds to [0 – 1]. A maximum at 30 °C (Euler beta function fit). Calculated data points and error bars – see SI Table 6. The fit parameters calculated from Eq. (2) are indicated in the chart. The goodness of fit: $\chi^2 = 0.00065$ and the determination coefficient $R^2 = 0.86$. The use of beta distribution (Euler beta function fit) has already been used by Yan and Hunt (1999) for temperature dependent plant growth.

**Figure 2 A** The effective diffusion rate $k_2$ of endogenous growth calculated from the growth functional, Eq. (1), for coleoptiles of maize grown at different temperatures (solid squares), and fitted to the Gauss function (solid line). **(a)** Interpolated by Microcal Origin: $T_1 = 14.6\,°C$ and $T_2 = 38.16\,°C$; half-widths $w_1 = 0.12 \pm 0.03$ and $w_2 = 0.73 \pm 0.49$. The values of the diffusion rate $k_2$ at low and high temperatures are diminished due to metabolic changes (see the text). The calculated data points and error bars – see SI Table 4. Fit parameters: $\chi^2 = 0.09056$ and determination coefficient $R^2 = 0.95$. Excluded area at temperatures exceeding 40 °C: see dense (red) pattern due to high error values (SI Table 4). Arrow pointing at the upper local maximum at $T_2$. **(b)** Interpolated by DAVE (Azuah et al. 2009): $T_1 = 15\,°C$ and $T_2 = 34.9\,°C$. In the suboptimal temperature range ($T < T_1$) plasmalemma is supposed to be in crystalline phase, in the optimal range ($T_1 < T < T_2$) – in semi-liquid phase, while in the supraoptimal range ($T > T_2$) – in liquid phase. In Figs 2A – C sub- and supra-optimum temperatures are excluded for fundamental (biochemical) reasons, which is apparently reflected in the raising errors obtained in the computer code (SI Table 4 – 6). Rescaling of abscissa in Figs 2A – C: [0 – 40] °C → [0 – 1]. **B** The effective diffusion rate $k_2$ for exogenous IAA calculated from growth functional, Eq. (1), for coleoptiles of maize grown at different temperatures (solid squares). Calculated data points and error bars from SI Table 5. The horizontal (blue) line represents – for optical reference only – a constant diffusion rate for $k_2 = 1.86 \cdot 10^{-4}\,s^{-1}$ for [8 – 40] °C interval. Excluded areas at temperatures below 8 °C and exceeding 40 °C: sparse and dense pattern, respectively, due to high error values (SI Table 5). A local minimum at about 30 °C appears (arrow). The diffusion rate starts

ascending above this value reaching a maximum at 40 °C. Then the diffusion rate drops down. Alternatively, the inset shows the linear interpolation for the non-excluded data. **C** The effective diffusion rate $k_2$ for exogenous FC calculated from growth functional, Eq. (1), for coleoptiles of maize grown at different temperatures (solid squares). Calculated data points and error bars from SI Table 6. Solid (red) line represents a linearly ascending diffusion rate for [0 – 40] °C interval. Linear model: slope (a) and intercept (b) are shown in the legend. Indicated confidence bands (dashed lines) at a confidence level α = 0.05. The local minimum (20 °C) and the local maximum (35 °C) are pointed by the arrows. The latter may be connected with the respective "phase transitions" taking place in the cell wall (see text). Excluded areas at temperatures exceeding 40 °C: dense pattern due to the calculated by the computer code high error values (SI Table 6).

**Figure 3** Cross-correlations of pH and elongation growth as a function of time lag τ, Eq. (5), for APW (endogenous growth), exogenous IAA and FC. Analysis based on the raw data presented in Figs 1 – 5 in Karcz and Burdach (2007). The time retardation (advancement) of the maximum with respect to zero time delay is clearly visible. Note a (red line) plot for APW at 40 °C which resembles an ideal cross-correlation of a step function and a rectangular triangle: the meaning is that growth and acidification takes place simultaneously (without retardation), which may be caused by the amplified diffusion at high ($T_2$) temperature (Fig. 1A). Maxima located at positive abscissa values correspond to the retardation of the medium pH.

**Figure 4** Cross-correlations of pH and elongation growth as a function of time lag τ parameterized by temperature, for APW (endogenous growth), exogenous IAA and FC. Data as in Fig. 3. The time retardation (advancement) of the maximum is clearly visible. The prevailing role of FC at sub- and supra-optimal temperatures is evident. Maxima located at positive abscissa values correspond to the medium pH retardation.

**Figure 5** Relative buffer acidification due to proton efflux. Cross-correlations derivative of elongation growth and pH as a function of time lag τ, Eq. (6), parameterized by temperature, for APW (endogenous growth), exogenous IAA and FC. Data as in Fig. 4. Discontinuities (jumps) in the cross-correlation derivative at τ = 0 correspond approximately to the (logarithm of) $H^+$ ions activity for maize coleoptile segments – numerical values are indicated in the legends.

**Figure 6** Relative buffer acidification due to proton efflux. Cross-correlations derivative of elongation growth and pH as a function of time lag τ (min) parameterized by temperature, for APW (endogenous growth), IAA and FC. Analysis based on raw data, Figs 1 – 5 in Karcz and Burdach (2007). Discontinuities (jumps) in the cross-correlation derivative correspond approximately to the (logarithm of) $H^+$ ions activity – numerical values are indicated in the legends.

**Figure 7** Cross-correlations, Eq. (5), calculated for the kinetics of (a) IAA-induced growth rate and proton efflux rate (solid dots) and (b) action of FC on growth rate and proton secretion rate (solid triangles) for the simultaneously measured both parameters, presented in Figs 4 – 5 by Lüthen et al.

(1990). Note that the timings for each plot almost coincide and the cross-correlation intensity for FC is about 4 times stronger than for IAA.

**Figure 8** Auto-correlations, $f = g$ in Eq. (5), calculated for the kinetics of (a) IAA-induced growth rate (**A**) and proton efflux rate (**B**) and (b) action of FC on growth rate (**C**) and proton secretion rate (**D**) for the simultaneously measured both parameters, presented in Figs 4 – 5 by Lüthen et al. (1990). Note, that auto-correlations deliver almost identical results (vertical scale neglected), both for fusicoccin (FC) and auxin (IAA) action. This result can be treated as convincing argument for the "acid growth hypothesis", applicable not only for FC but IAA as well. Growth rate and proton efflux rate coincide for both FC and IAA, and can be used interchangeably.

**Figure 9** Cross-correlation, Eq. (5), calculated for the simultaneous measurement of the medium pH of 52 abraded maize coleoptile segments in 5 mL of aerated medium and the relative growth rate of a representative stack of four of them presented in Fig. 3 by Peters et al. (1998). Curve maximum corresponding to the maximum of growth rate after addition of IAA.

**Figure 10** Cross-correlations, Eq. (5), calculated for the kinetics of (**A**) IAA-induced growth rate and proton efflux rate, and (**B**) action of FC on growth rate and proton secretion rate for the simultaneously measured (by the authors) both parameters (SI Fig. 5) on intact growing maize seedlings for different, but identical for IAA and FC concentrations (control means APW). The retardation (shift of a maximum) clearly visible for FC may be due to the lower proton diffusion rate to the buffer for the 'intact seed' experiment.

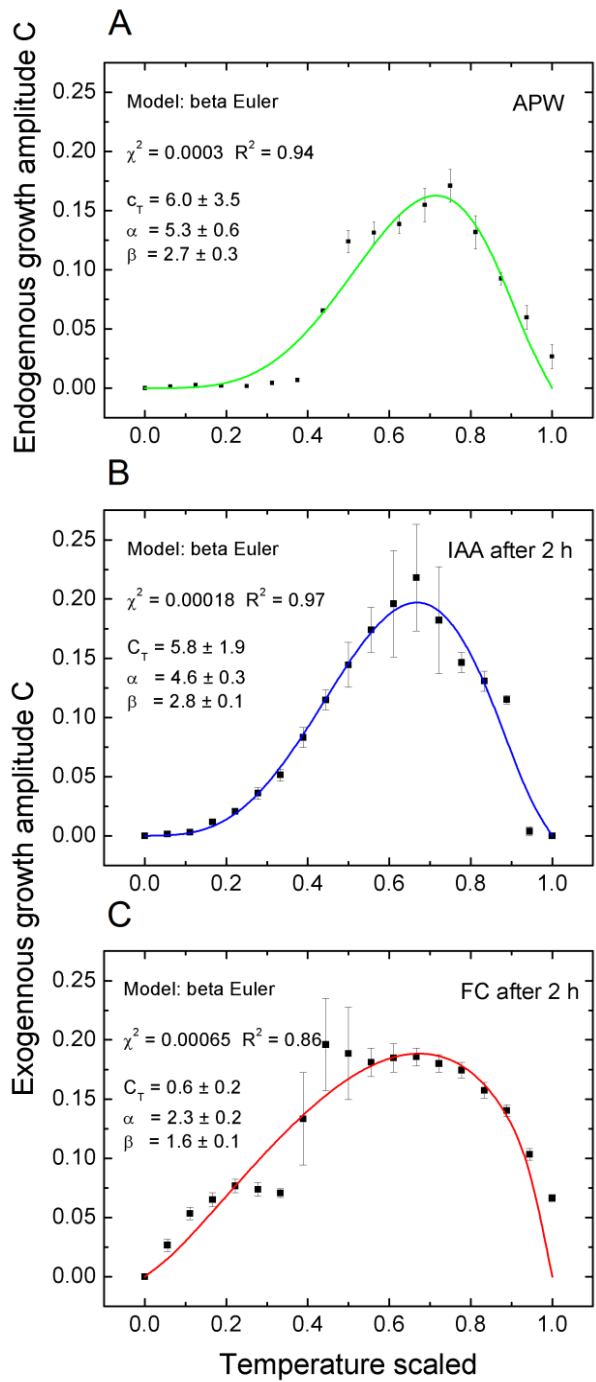

**Figure 1**

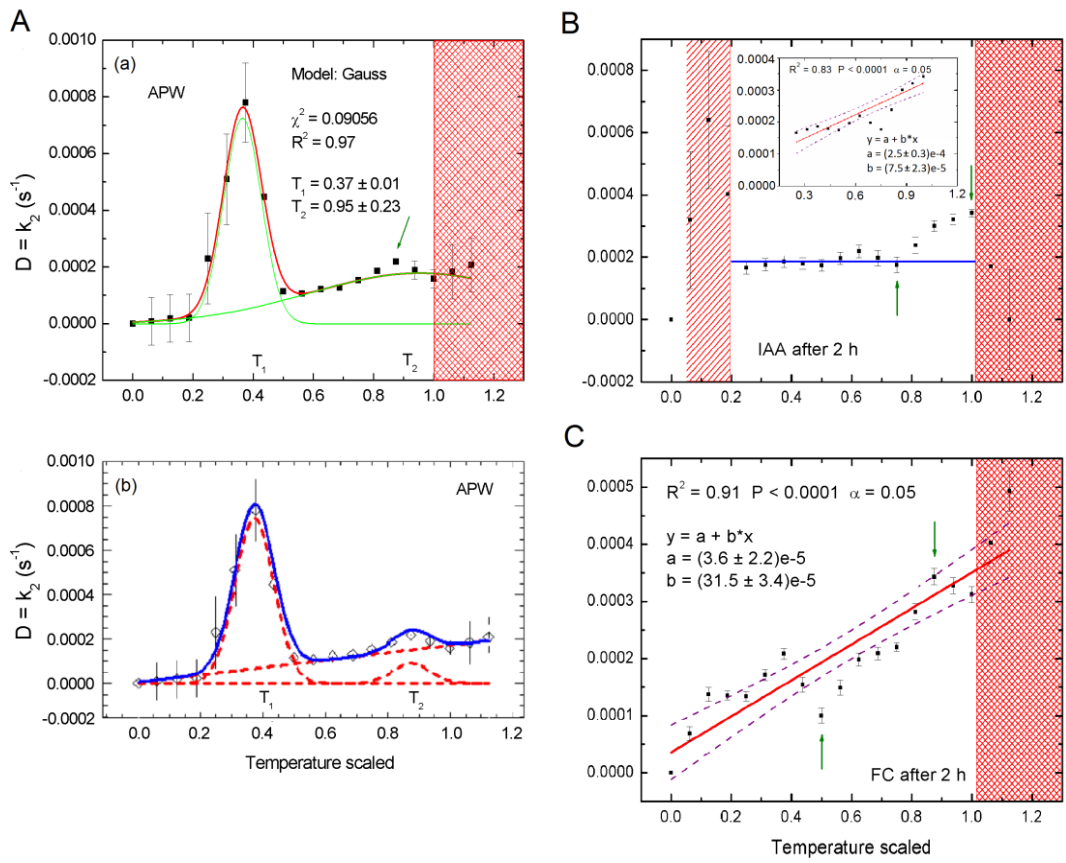

**Figure 2**

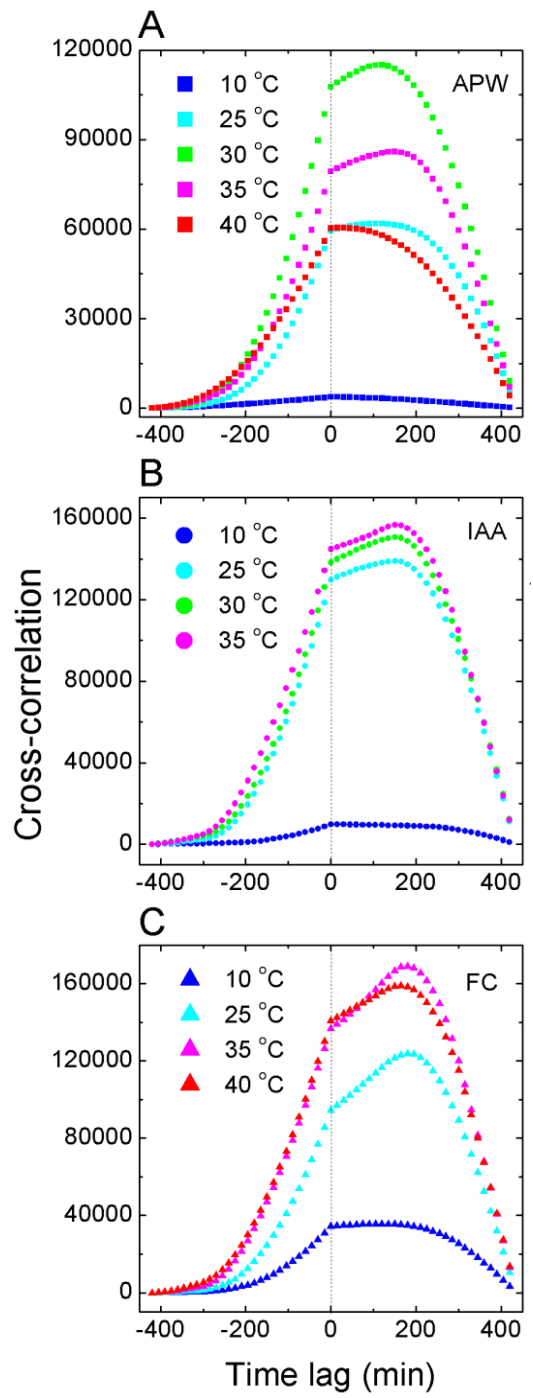

**Figure 3**

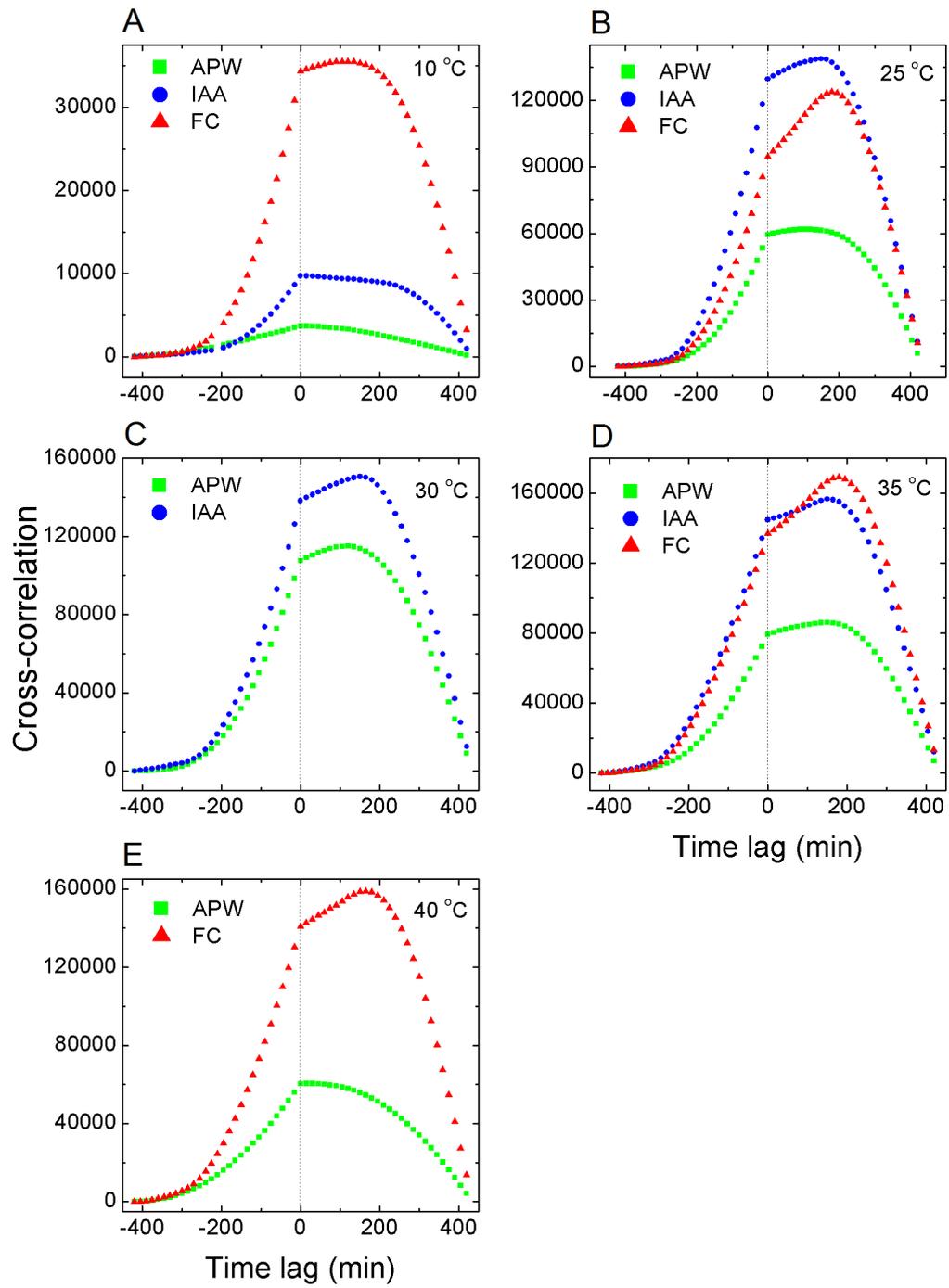

**Figure 4**

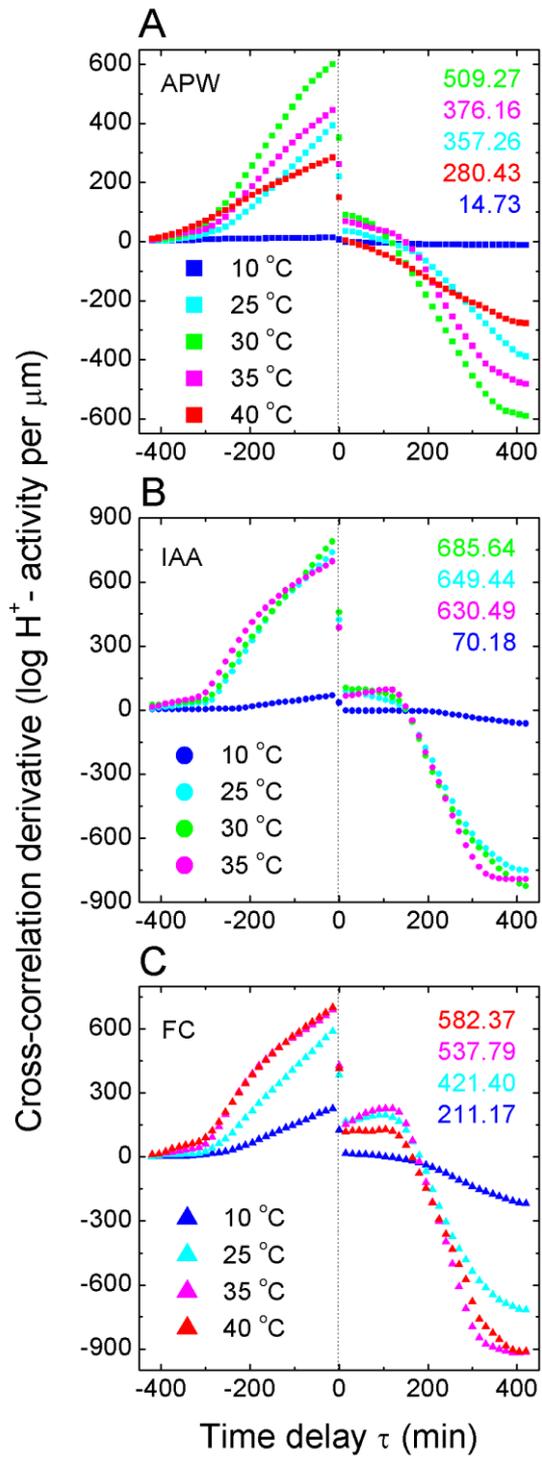

**Figure 5**

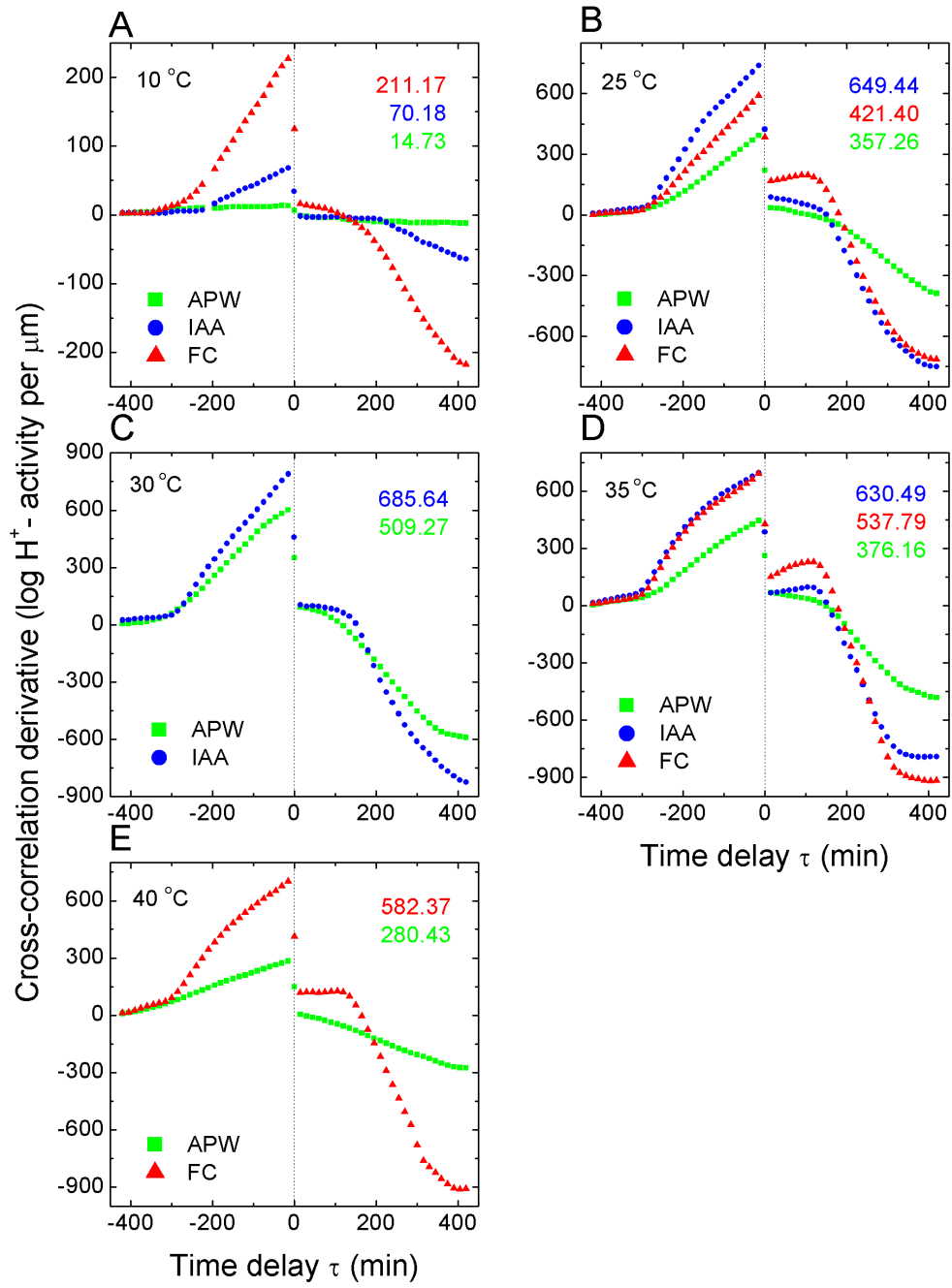

**Figure 6**

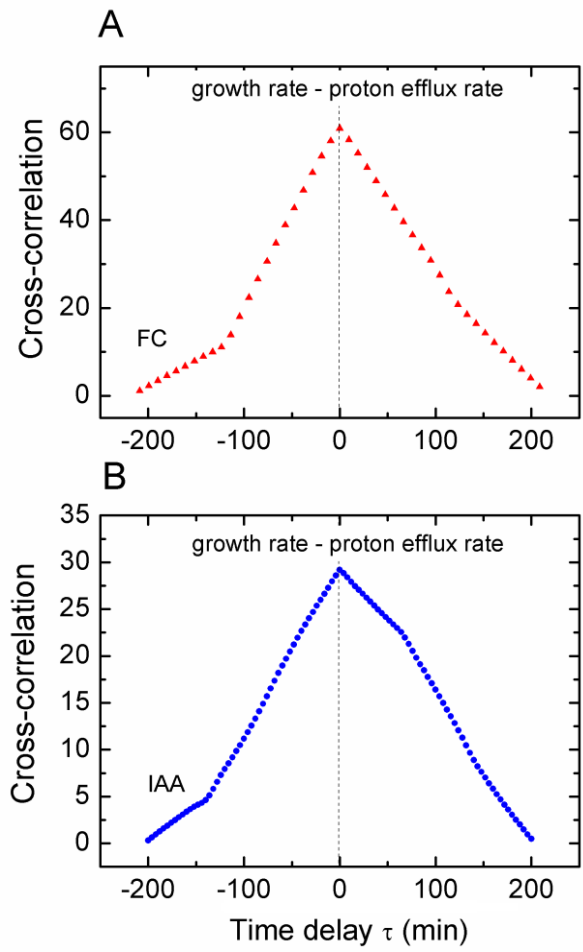

**Figure 7**

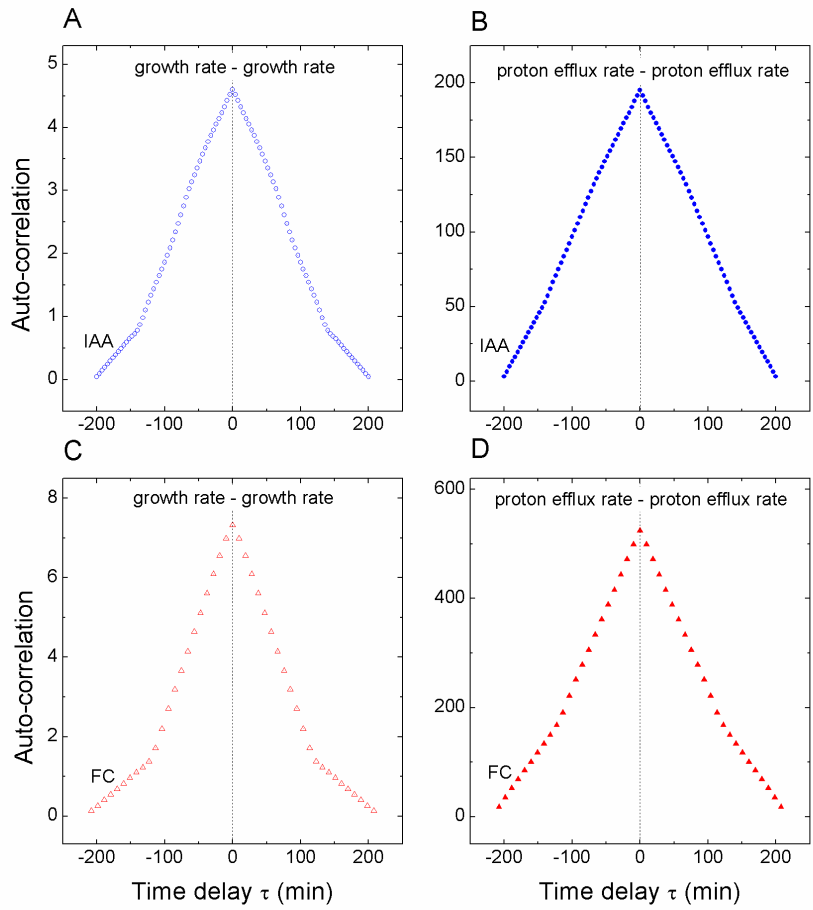

**Figure 8**

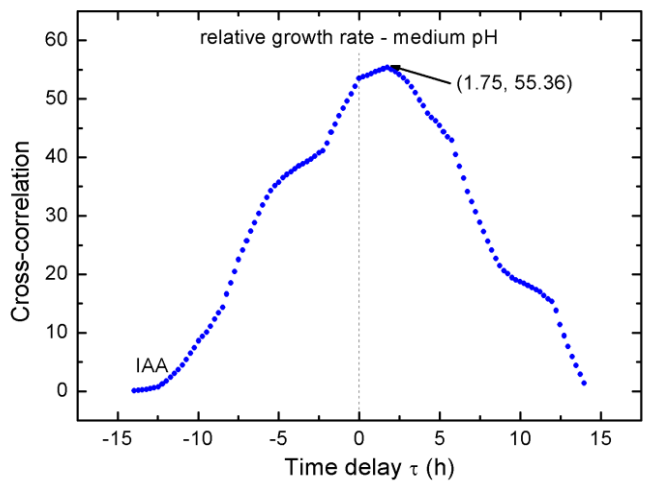

**Figure 9**

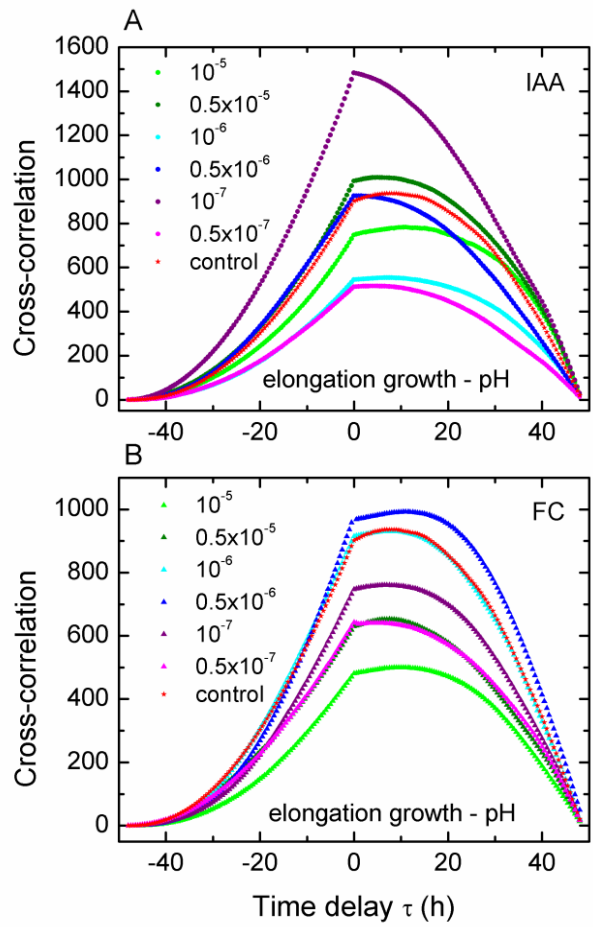

**Figure 10**